\documentclass{aa}
\usepackage{graphicx}
\begin{document}

\newcommand{\kms}{km~s$^{-1}$}
\newcommand{\cm}{cm$^{-2}$}
\newcommand{\lya}{Lyman~$\alpha$}
\newcommand{\lyb}{Lyman~$\beta$}
\newcommand{\za}{$z_{\rm abs}$}
\newcommand{\ze}{$z_{\rm em}$}
\newcommand{\nhi}{$N$(H~I)}
\newcommand{\zae}{$z_{\rm abs} \sim z_{\rm em}$}

\def\ltsima{$\; \buildrel < \over \sim \;$}
\def\simlt{\lower.5ex\hbox{\ltsima}}
\def\gtsima{$\; \buildrel > \over \sim \;$}
\def\simgt{\lower.5ex\hbox{\gtsima}}
\def\arcs{$''~$}
\def\arcm{$'~$}

   \title{The CORALS Survey II: Clues to Galaxy Clustering Around
QSOs from \zae\ Damped Lyman Alpha Systems.\thanks{The work presented 
here is based, in part, on data obtained with the ESO facilities on 
La Silla (EFOSC/3.6-m) and Paranal (FORS1/UT1), programs 66.A-0037 and 
267.A-5686. }
}

 \titlerunning{CORALS II: \zae\ DLAs}

   \author{S. L. Ellison\inst{1}
          \and
          L. Yan\inst{2}
          \and
          I. M. Hook\inst{3}
        \and
        M. Pettini\inst{4}
        \and
        J. V. Wall\inst{5}
        \and
        P. Shaver\inst{6}
          }

   \offprints{S. Ellison}

  \institute{European Southern Observatory, Casilla 19001, Santiago 19, 
        Chile\\
              \email{sellison@eso.org}
        \and
	SIRTF Science Center, Caltech, California, USA\\
	 \email{lyan@ipac.caltech.edu}
	\and
	Astrophysics: Department of Physics, Nuclear and Astrophysics 
        Laboratory, Keble Road, Oxford,  OX1 3RH, UK\\
	\email{ihook@gemini.edu}
	\and
	Institute of Astronomy, Madingley Rd., Cambridge, CB3 0HA, UK\\
	\email{pettini@ast.cam.ac.uk}
	\and
	Astrophysics: Department of Physics, Nuclear and Astrophysics 
        Laboratory, Keble Road, Oxford,  OX1 3RH, UK\\
	\email{jvw@astro.ox.ac.uk}
	\and
	European Southern Observatory, Karl-Schwarzschild-Str. 2, D-85748
        Garching bei Munchen, Germany
	\email{pshaver@eso.org}
}


   \date{Received / Accepted}

\abstract{
We present a list of \zae\ ($\Delta v < 3000$\,\kms)
DLAs discovered during the CORALS survey
for absorbers in a radio-selected QSO sample.  On the assumption that
\zae\ DLAs are neither ejecta from the QSO, nor part of the host galaxy
itself, we use the number density statistics of these DLAs
to investigate galaxy clustering near the QSO redshift.
We find that $n(z)$ of \zae\ DLAs in our radio-selected QSO 
sample is $\sim 4$ times larger
than the number density of intervening DLAs, implying an excess
of galaxies near the QSO.  This result is further supported with
the inclusion of the radio-loud QSOs in the FIRST survey, although the
total number of DLAs is still small (4) and the result is currently
only significant at approximately the $2\sigma$ level. 
Since all of the \zae\ DLAs we identify in CORALS
are found towards optically bright ($B < 20$) QSOs, there is no
strong evidence (based on these limited statistics) 
that this population suffers from a severe dust bias.
We compare our results with those from an optically-selected, radio-quiet 
QSO sample in order to determine whether there is evidence for an excess
of galaxies around radio-loud versus radio-quiet QSOs. We find that
the $n(z)$ of \zae\ DLAs towards radio-quiet QSOs is in agreement with
the number density of intervening absorbers.  This result, although
currently limited by the small number statistics of our survey, supports
the conclusion that radio-loud quasars are found 
preferentially in rich galaxy environments at high redshift.
Finally, we report that one of the new \zae\ DLAs
discovered by CORALS has some residual flux in the base of the
\lya\ trough which may be due to \lya\ emission, either from star
formation in the DLA galaxy
or from gas photoionised by the QSO.
   \keywords{Quasars: general -- quasars: absorption lines -- 
galaxies: evolution -- galaxies: clusters: general}
}

\maketitle
%

\section{Introduction}

The formation of quasars is believed to represent
an integral stage in 
the process of galaxy formation and evolution.  
The high incidence of
galaxy-QSO companions and apparent mergers supports the view that 
quasar activity is fuelled by continued accretion of material onto 
the central black hole from the intercluster medium or 
nearby galaxies (see the review by Barnes \& Hernquist 1992).
Many of the outstanding questions in theories of the role of
AGN in galaxy evolution involve the link between 
radio power and the properties of the QSO environment.  
The long-standing view is that both the morphology of the
host galaxy (e.g. Hamilton, Casertano \& Turnshek 2001),
and the richness of the surrounding environment
(Yee \& Green 1987; Hall \& Green 1998; Hall, 
Green \& Cohen 1998; Hutchings et al. 1999; 
Teplitz, McLean \& Malkan 1999; Cimatti et al. 2000) 
are different between radio-quiet QSOs (RQQs) and radio-loud 
QSOs (RLQs). 
Specifically, Yee \& Green (1987) pointed out that the former
are rarely found in clusters, while 
about 35\% of RLQs show an excess of faint galaxies 
in their vicinity, consistent with 
the presence of a cluster of galaxies at the QSO redshift
(see also Sanchez \& Gonzalez-Serrano 1999).  Radio galaxies,
which are intimately connected with RLQs in unification models,
also appear to reside in cluster environments at high redshift
(Pentericci et al. 2000).
However, the dichotomy between radio-loud and radio-quiet properties 
is now being questioned by studies which have found that almost all QSO 
hosts are elliptical galaxies, regardless of 
radio-loudness (McLure et al. 1999; Dunlop et al. 2002), and that the
environments of RLQs are statistically indistinguishable from
those of RQQs, at least at $z \simlt 1$ 
(Hutchings, Crampton \& Johnson 1995; McLure \& Dunlop 2001; 
Wold et al. 2001; Finn, Impey \& Hooper 2001).  At higher
redshifts, there is evidence that galaxy density maybe higher
around RLQs, although a milder overdensity is seen towards
RQQs (see review by Hutchings 2001).
This ongoing discussion is made even more relevant 
by (a) our lack of understanding
of the mechanism responsible for the radio power of RLQs
and (b) the
recent paradigm shift away from the long held belief 
that the distribution of radio luminosities is bimodal (Kellermann et al 1989;
White et al. 2000; Brinkmann et al. 2000;  Lacy et al. 2001). 
It is of primary importance to establish the extent to
which radio power depends on the QSO environment, since this
relationship clearly 
holds important clues to the physics of AGN and 
the role of quasars in galaxy formation.  

In this paper we explore the possibility of using quasar absorption
lines as an unbiased probe of the high redshift QSO environment.
This is a potentially powerful technique since studies of
galaxy clustering around QSOs can be extended to much higher
redshifts, although the obvious disadvantage is the uni-dimensional
information content.  Therefore, a large number of QSOs needs to be
studied in order for the statistical trends of \zae\ absorbers
to provide meaningful results.  In this respect, the results 
presented here for the 66 QSOs of the Complete Optical and Radio Absorption
Line System survey (CORALS, Ellison et
al. 2002) should be regarded as preliminary, although
they underline the potential of this approach, so far relatively unexploited.
 
There is already some evidence that the posited correlation between 
galaxy concentration and radio-loudness may extend to 
QSO absorption line systems,
based on an excess of associated C~IV systems
(Foltz et al. 1986; Anderson et al. 1987; 
Foltz et al. 1988) in RLQ spectra
out to 3000 \kms\ from the QSO redshift.   
However, the origin of these absorbers is unclear. The excess seems to be 
restricted to steep-spectrum RLQs, 
and may thus be related to some orientation
effect (Barthel et al. 1997; Richards et al. 1999; 
Richards 2001). On the other hand,
Richards et al. (2001) find no correlation
of this excess with other orientation measures
such as core-to-lobe ratio.  
Nevertheless, the suspicion remains
that these unusually strong and highly
ionised systems are probably
intrinsic to the QSO and may not be  
providing us with specific clues about the surrounding
galaxy environment (Baker et al 2001).
A more fruitful line of investigation is to focus on
\zae\ absorbers which are analogous to galaxy-scale absorbers
at intervening redshifts, namely the Damped Lyman Alpha systems (DLAs).

In the last 15 years, several surveys have mined the sky for DLAs, 
with the objective of understanding more 
about the high redshift galaxy population that these
absorption systems are thought to represent (Wolfe et al. 1986;
Lanzetta et al. 1991, 1995; Wolfe et al. 1995; Storrie-Lombardi \&
Wolfe 2000; P\'{e}roux et al. 2001).  Almost without exception, these
surveys have adopted the early strategy and definitions of Wolfe 
et al. (1986) who included in their sample 
only DLAs with $N$(H~I) $\ge 2 \times
10^{20}$ atoms \cm\ that lie at absorption redshifts corresponding to
$\Delta v >$ 3000 \kms\ from the QSO emission redshift.  The reason
for imposing this velocity cut is primarily due to the unknown
nature of these proximate DLAs (PDLAs), which may conceivably be caused
by QSO ejecta or by absorption in the host galaxy itself.  In addition,
since one of the main objectives of DLA surveys has been to compile
a census of H~I in the universe, proximate systems have been excluded 
due to the possible effect of the QSO's ionizing radiation on the absorber.  
For example, the intense local radiation field of the QSO is already 
well known to affect the distribution of the lower column density \lya\ 
forest clouds (e.g. Murdoch et al. 1986).  A similar phenomenon has
also been noted to affect the extended \lya\ haloes of low redshift
galaxies (Pascarelle et al. 2001).  Therefore,
in the quest to gain an insight into the `normal and representative'
galaxy population at high redshift, PDLAs at \zae\ have been almost 
universally discarded from the statistics of previous surveys.  

M\o ller, Warren \& Fynbo (1998),
motivated by observations of \lya\ emission in associated DLAs,
have explored the possible nature of these systems.  They concluded 
that PDLAs are unlikely to be due to QSO ejecta,
based on their lower metallicities, lack of strong high ionization lines
and simpler line profiles compared to intrinsic BAL features (see
Barlow et al. 1997 for a summary of intrinsic absorber signatures).
Although recent models for the structure of AGN (e.g. Elvis 2000)
include relatively high column density, wind-ejected clouds, the
consequent UV absorbers are quite different from 
PDLAs both in terms of $N$(H~I) and ejection velocity
(e.g. Monier et al. 2001).  For example, narrow intrinsic C~IV systems, 
identified by means of time variability or partial coverage
have relatively low H~I column
densities (e.g. Petitjean, Rauch \& Carswell 1994; 
Hamann, Barlow, \& Junkkarinen 1997).  It is also unlikely that PDLAs 
are generally due to the QSO host galaxy, again because of the significant
velocity differences involved.   
It therefore seems highly
plausible that PDLAs are either members of the same family as,
or at least strongly resemble, the population of intervening absorbers.
If this interpretation holds true, then investigating the hitherto
unexploited sample of PDLAs may yield further insights into the
connection between AGN activity and galaxy formation.

Based on a radio-selected sample of QSOs, 
CORALS (Ellison et al. 2002)
is one of the most recent DLA surveys aimed at
determining the extent of any dust bias affecting the statistics
of DLAs drawn from 
optically-selected, magnitude limited QSO samples.
In this paper, we present the PDLAs found by the CORALS survey (\S2)
and compare their rate of incidence with that of intervening absorbers
at $\Delta v > 3000$\,km~s$^{-1}$ in the same sample of QSOs (\S3).
We also compare the frequency of PDLAs between RLQs and RQQs.
Our main result is that we find a possible excess of PDLAs towards RLQs
(currently only confirmed at slightly below the $2\sigma$ level),
and that since all of the systems that we identify lie in front of optically
bright QSOs it appears that PDLAs contain only a modest amount of dust. 
These findings, which are still tentative
because of the small size of present samples,
are briefly discussed in \S4.

\section{Identified PDLAs in CORALS}
 
\begin{figure}
\centerline{\rotatebox{0}{\resizebox{8.8cm}{!}
{\includegraphics{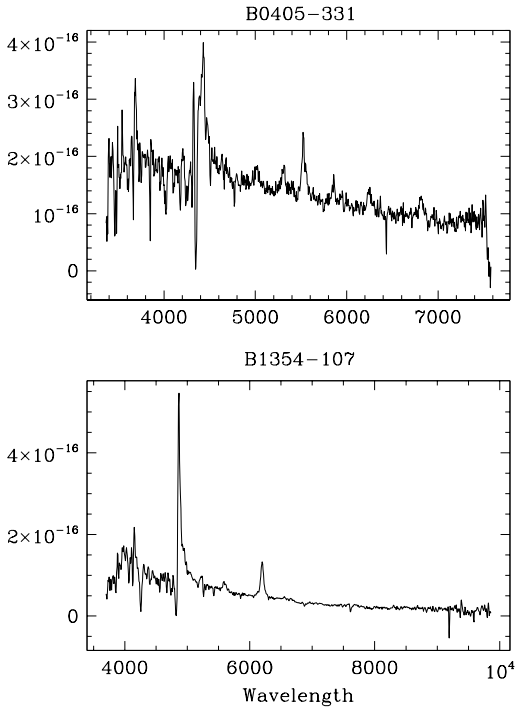} }}}
\caption{\label{disc_spec} Discovery spectra (obtained with EFOSC at
the 3.6-m on La Silla) of two of the CORALS QSOs with proximate DLAs
showing that these sources are not BAL QSOs. A third QSO
with a PDLA, B0528$-$250, was already known and was not reobserved.}
\end{figure}

Full details of the CORALS sample definition, observations and all spectra
can be found in Ellison et al. (2002).  In brief, a sample of 66 
\ze\ $\geq 2.2$ radio-loud QSOs was selected 
from the flat-spectrum ($\alpha > -0.4$) sources in PKSCat90, 
the catalogue compendium from the Parkes surveys. A full list
of flux densities, source positions and spectral indices is given
Jackson et al. (2002).  For this
radio-selected sample, an optical spectrum was obtained for
each QSO in order to search for DLAs with $1.8 <$ \za\ $<$ \ze.
Out of a total of 22 confirmed DLAs, three were at redshifts corresponding
to velocities $\Delta v \leq 3000$\,\kms\ (the original limit
adopted by Wolfe et al. 1986).  We note that although the \lya\
line used here to determine \ze\ may be systematically blue-shifted
with respect to other emission lines, implying that the actual velocity
offset of the PDLAs maybe be up to $\sim$ 4500 \kms\ from \ze.  However,
this does not affect our analysis because the 3000 \kms\ limit
traditionally adopted to distinguish intervening systems is measured
\textit{relative to Ly$\alpha$}, not H$\alpha$ or any highly ionized lines.
The three PDLAs are in the sightlines
towards B1354$-$107b, B0405$-$331 and B0528$-$250.  The third of these 
systems has been known for a long time -- Morton et al. (1980) measured
$z_{\rm abs} = 2.811$ and \nhi\ = $1.6 \times 10^{21}$ \cm\ -- 
and was therefore not re-observed as part of the 
CORALS survey.  In Figure \ref{disc_spec} we present the original
low resolution spectra of the other 2 QSOs found to exhibit
PDLAs, obtained 
to confirm the QSO identification and redshift.  The large wavelength 
coverage out to beyond C~IV allows us to confirm that these are not BAL QSOs
and that the PDLAs are  therefore unlikely to be high velocity ejected
material.  Higher resolution (FWHM $\sim$ 3 \AA) spectra have been obtained of
these targets with the AAT, see Ellison et al. (2002) for spectra and
details of observation and data reduction.  Figure 2 shows close-ups 
of the damped \lya\ lines together with adopted fits to the line profiles.
 
\begin{figure}
\centerline{\rotatebox{0}{\resizebox{8.8cm}{!}
{\includegraphics{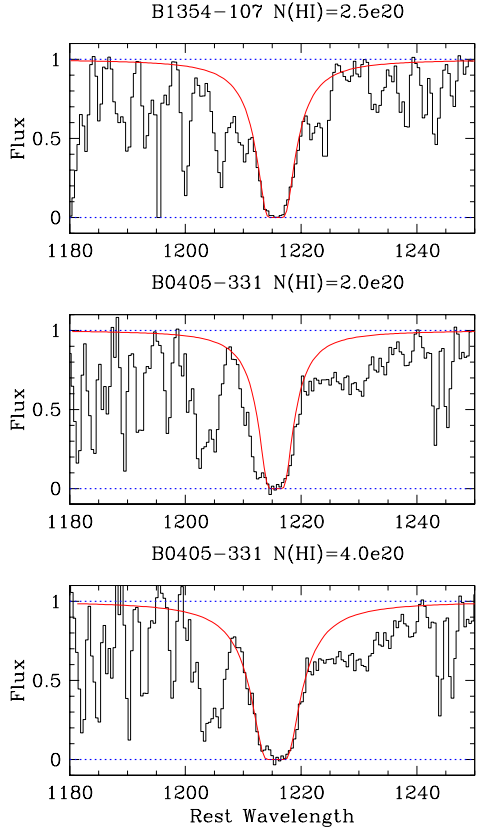}}}}
\caption{\label{dla_fit}
H~I fits to the 2 new PDLAs found in the CORALS
sample.  Top panel:  Fit to the $z_{\rm abs} = 2.966$ DLA towards B1354$-$107.
Middle panel:  Possible fit to the DLA towards B0405$-$331 with
$z_{\rm abs} = 2.572$ and \nhi\ = $2 \times 10^{20}$\cm.  Bottom panel:
Alternative fit to the DLA towards Q0405$-$331 with $z_{\rm abs} = 2.570$ 
and $N$(H~I) = 4$ \times 10 ^{20}$ \cm.  In this case, some residual
flux can be seen at the blue edge of the \lya\ trough.  }
\end{figure}

We will now briefly describe the properties of these two
new systems. 

\paragraph{B1354$-$107b}  Two DLAs are detected in the sightline
towards this QSO, the higher redshift absorber `b' 
having a redshift of $z_{\rm abs} = 2.966$.  The velocity difference 
between the absorber and the emission redshifts ($z_{\rm em} = 3.006$) is
therefore $\sim$ 3000 \kms, just on the limit of our definition of
a PDLA given the uncertainty of \ze\ and \za.  The Lyman limit at $\lambda_0$ =
912\AA\ associated with this system is clearly visible at
$\lambda_{obs} \sim 3650$ \AA\ in our AAT spectrum (see Ellison et al.
2002).  The H~I fit to this absorber is
complicated by the onset of \lya\ emission in the absorber's red
wing.  No information about the column density can therefore be
gleaned from this part of the line profile.  There is some additional,
blended absorption in the blue, so the fit is mainly constrained by
the base of the trough, see Figure \ref{dla_fit}.  

\paragraph{B0405$-$331}  This DLA has an absorption redshift
$z_{\rm abs} \sim 2.57$ which is the same as the QSO redshift
of  $z_{\rm em}$ = 2.570.  The absorption trough of this system is
superimposed on the \lya\ emission from the quasar itself, rendering an
accurate continuum fit very difficult. 
Figure \ref{dla_fit} (middle panel)
shows that a good fit to the damped \lya\ line
is obtained  with a $N$(H~I) = $2 \times
10^{20}$ \cm\ at $z_{\rm abs} = 2.572$.\footnote{Although this is 
nominally higher than the emission redshift, the difference 
is only 168\,\kms, well within the uncertainty in the QSO systemic
redshift and within the velocity dispersion of a galaxy cluster
of which the QSO host galaxy may be a member.} 
However, we find that an absorber at
$z_{\rm abs} = 2.570$ and with
$N$(H~I) = 4$ \times 10 ^{20}$ \cm\ would also
produce a good fit to the base and the blue slope of the absorption
profile, although there is a small but significant residual flux
feature visible in the wing of the trough (bottom panel of Figure 
\ref{dla_fit}).  Several associated metal lines are present in our spectrum,
particularly 
C~II $\lambda$1334, Si~II $\lambda$1260 and Si~II
$\lambda$1190 with redshifts of $z_{\rm abs} = 2.569$, 
2.569 and 2.570 respectively.
\lyb\ is also covered by these observations and is found at a redshift
of 2.570.  These lines therefore more closely match the lower
redshift, higher column density solution shown in the bottom panel of
Figure \ref{dla_fit}.   This would then leave 
a small residual flux in the base of the DLA absorption 
which we interpret as \lya\ emission
associated with the absorbing galaxy.
Its velocity offest from the  
centre of the absorption trough is $\Delta v = -300$\,\kms,
very similar to that between the 
DLA and \lya\ emitter in Q0151+048A (M\o ller, Warren
\& Fynbo 1998; Fynbo, M\o ller \& Warren 1999).

There is no obvious CCD defect or
cosmic ray event at the position of the residual flux feature on the 2D
spectrum. This is clearly a system which should be followed 
up with narrow band imaging as it would 
add to the handful of known DLAs which have associated
\lya\ emission.  It is interesting to 
note that there is some evidence that PDLAs are more 
likely to exhibit strong \lya\ emission than $z_{\rm abs} \ll z_{\rm em}$
`intervening' DLAs (M\o ller, Warren \& Fynbo 1998).  
Although the presence of \lya\ emission may in some cases be caused by 
photo-ionization (e.g. Fynbo, M\o ller \& Warren 1999; 
Warren et al. 2001), there
is also evidence that star formation may be involved in 
other cases (e.g. M\o ller \& Warren 1996; Ellison et al. 2001).

\section{Results}

\subsection{The Number Density of PDLAs in the Radio-Loud CORALS Sample}

\begin{table*}
\begin{center}
\begin{tabular}{lcccccccc}\hline \hline
 & & & & & & &\\
Sample & No. QSOs & QSO Type & No. DLAs & $\Delta z_{3000}$ & $n(z)$ PDLA& 
$\langle z_{abs} \rangle$ & $n(z)$ DLA\\
 & & & & & & &\\ \hline
 & & & & & & &\\
CORALS & 66 & RLQ & 3 & 2.48 & 1.21$^{+1.17}_{-0.65}$ & 2.78 & 0.24\\
CORALS+FBQS & 96 & RLQ  & 4 & 3.58 & $1.12^{+0.88}_{-0.54}$ & 2.76 & 0.24\\
P\'{e}roux et al. & 49 & RQQ & 1 &2.56 & 0.39$^{+0.35}_{-0.31}$ & 4.26 & 0.35\\
 & & & & & & &\\ \hline
\end{tabular}
\caption{\label{n_stats} Number density statistics for PDLA samples in 
CORALS, the FBQS survey (White et al. 2000) and the survey of P\'{e}roux 
et al. (2001).  The last column gives the expected number density of PDLAs
according to the statistics for intervening DLAs
determined by Storrie-Lombardi \& Wolfe (2000).}
\end{center}
\end{table*}

The statistics of our survey are collected in Table 1.
The number of damped absorbers found per unit redshift can be
expressed as the DLA number density, $n(z)$, quoted for a specific mean
absorption redshift.  This is simply the total
number of DLAs identified in a given sample, divided by the total
intervening redshift interval covered.  The redshift interval, $\Delta
z$, is given by
 
\begin{equation}
\Delta z = \sum_{i = 1}^{n} (z_{i}^{max} - z_{i}^{min})
\end{equation}

where the summation is over the $n$ QSOs in a given sample.
In the case of PDLAs, we only include the redshift interval from
the emission redshift extending out to 3000 \kms,
 
\begin{equation}
\Delta z_{3000} = \sum_{i = 1}^{n} (z_{i}^{em} - z_{i}^{3000})
\end{equation}

where $z_{i}^{3000}$ is the redshift corresponding to a relative 
velocity of 3000\,\kms.  

For the CORALS survey, we find $n(z) = 1.2^{+1.2}_{-0.7}$ at a mean
$\langle z \rangle = 2.78$
(using $1 \sigma$ confidence limits appropriate
to the Poissonian small number statistics 
formulation by Gehrels 1986). 
This value is $4^{+4}_{-2.4}$ times larger than 
value $n(z) = 0.31^{+0.09}_{-0.08}$ 
at $\langle z \rangle = 2.37$ determined for
the intervening DLAs in the CORALS sample by Ellison et al. (2002).  
We find a similar apparent excess compared to the 
density of intervening DLAs determined from optical samples.
Adopting $n(z) = 0.055(1 + z)^{1.11}$, as deduced 
by Storrie-Lombardi \& Wolfe (2000), we would 
have expected only $n(z) = 0.24$ at $\langle z \rangle = 2.78$.
However, we are dealing with very small number statistics here.
Since the redshift path $\Delta z$ which can be probed for 
PDLAs in a given QSO sight-line is very small, a much larger sample
is required to establish conclusively if PDLAs are indeed more common
towards RLQs.

In an effort to improve the statistics, 
we also consider the recent radio-selected FIRST Bright
Quasar Survey (FBQS; White et al. 2000). 
Specifically, we selected the 41 QSOs with 
$z_{\rm em} \geq 2.2$ in this independent sample (excluding BALs).  
Due to the different selection criteria and
survey sensitivities of the FBQS compared to CORALS, we further exclude
the 11 FBQS QSOs that do not conform to the accepted definition of
radio-loud, i.e. with a radio-loudness parameter
$\log R^* \ge 1.0$ (Stocke et al. 1992).
Combining the remaining 30 FBQS targets (with 1 PDLA)
with CORALS, we cover a total 
interval $\Delta z_{3000} = 3.58$ and deduce $1.1^{+0.9}_{-0.5}$.
Thus, although the errors are still large, these statistics
indicate that 
\textit{there may be an excess of absorbers close to the systemic
redshift of radio-loud QSOs}.
If the postulate that PDLAs are not directly associated with the
QSO is true (M\o ller, Warren \& Fynbo 1998), then this excess may 
be analogous to the excess
of galaxies observed in the fields of RLQs (Yamada et al. 1997; Hall \&
Green 1998; Hall, Green \& Cohen 1998; Clements 2000).  Indeed, the
excess of DLAs close to the QSO compared with intervening systems is
consistent with the typical over-densities of galaxies found around RLQs by 
Sanchez \& Gonzalez-Serrano (1999) and of \lya\ emitters around radio
galaxies (Kurk et al. 2001).

\subsection{Comparison of the Number Density of DLAs between RLQs and RQQs}

If PDLAs are associated with galaxies clustering around QSOs,
then comparing the properties of these absorbers in radio-
versus optically-selected QSO sample will provide further evidence as
to whether RLQs preferentially flag rich cluster
environments at high redshift.  
However, this is not a simple undertaking for the following reason.
Most of the
early DLA surveys initially obtained low resolution (typically 10 \AA)
spectra of all the QSOs in a given sample. From this, a list of candidate
DLAs with rest frame equivalent width $W_0 \ge$ 5 \AA\ was compiled
 and subsequently followed up at higher
resolution to confirm the DLA nature and measure the column density
$N$(H~I).  However, in most cases PDLAs were
{\it not} included in the follow-up spectroscopy.
Ellison (2000) investigated how the probability
that an absorption system is damped varies as a function of 
$W_0$ and found that $\sim$ 90\% of DLA candidates with $W_0 > 10$ 
\AA\ were eventually confirmed to have $N$(H~I) $\ge 2
\times 10^{20}$ \cm.  Therefore, it is in principle possible
to use these statistical likelihoods to infer the number of 
PDLAs in the previous major surveys.  However, since 
PDLAs have been universally excluded from these surveys, even the
equivalent widths have generally not been tabulated in the literature.  
Given this unsatisfactory state of affairs, here
we restrict ourselves to comparing our estimate of $n(z)$ of 
PDLAs in CORALS with analogous statistics from 
the recent survey by P\'{e}roux et al. (2001).  

From the P\'{e}roux et al. (2001) sample of 66 QSOs, we exclude all targets 
which exhibit BAL features (10) or are confirmed to be radio-loud (7)
based on observations from NVSS and SUMSS measurements.
From the remaining 49 QSOs,
only 1 PDLA is found ($z_{abs} = 4.26$ towards PSS J0034+1639).  The
total redshift interval covered is $\Delta z_{3000}=2.56$ 
(P\'{e}roux, private communication), yielding a mean
$n(z) = 0.39$ at $\langle z \rangle = 4.26$, in good agreement
with Storrie-Lombardi \& Wolfe's (2000) derived $n(z)$ = 0.35 for
intervening DLAs at this redshift.  Note that despite the difference in
mean absorption redshifts between the RLQ and RQQ samples,
cluster evolution is unlikely to explain the disparate $n(z)$ due to
the small time ($\sim$ 0.5 Gyr) between these two epochs.
\footnote{We note that out of the 49 QSOs considered
in these statistics, 4 have no radio data available.  Since only 
approximately 10\% of QSOs are radio-loud,
we make the assumption that none of these 4 targets is radio-loud.
Even in the extreme and unlikely case that ALL of these 4 QSOs turned
out to be radio-loud, $\Delta z$ would be reduced by less than 10\%
to 2.35, i.e. $n(z) = 0.42$, a negligible change.}   
It therefore appears that \textit{the tentatively identified
overdensity of DLAs near to the QSO is confined to RLQs}.

\subsection{Dust Extinction in PDLAs}

Studies of optical (Outram et al. 2001) and IR
(Hewett, private communication) colours of large, optically-selected 
QSO samples find that the effect of dust reddening in $intervening$
absorbers is small.  Conversely, Carilli et al. (1998) found that 
a high fraction of `red' QSOs have associated \zae\ 21 cm absorption,
indicating that dust reddening may play a significant effect in proximate
absorbers.  Consequently, we may
expect that optically-selected, magnitude-limited QSO samples
may be biased against the detection of PDLAs.  It is
therefore somewhat surprising that all of the PDLAs of the CORALS
sample are found towards relatively bright QSOs ($B < 20$).
Taken at face value, this implies that any dust extinction which may
be present does not bias the statistics significantly,
although it must be remembered that we are dealing with very small number
statistics.  Since Ellison et al. (2002) find that up to
twice as many intervening DLAs are present towards faint QSOs, it will
be interesting to see whether observing more QSOs reveals a similar
effect for proximate DLAs.  
This would establish whether PDLAs are relatively dust-poor
or dust-rich.



\section{Conclusions and Discussion}

The main aim of this paper has been to assess whether PDLAs can provide 
an insight into the the clustering of galaxies around QSOs at high
redshifts.  

The main conclusions of this work may be summarised as follows.

\begin{enumerate}

\item  Despite the small number statistics of this work (4 DLAs in
the combined CORALS + FBQS sample), we find
evidence that our radio-selected sample of QSO exhibits
an excess (by a factor of 4) of PDLAs at \zae\ compared to
intervening redshifts.  This implies an excess of material 
(presumably galaxies) near the QSO at levels consistent with faint
galaxy excesses around RLQs (e.g. Sanchez \& Gonzalez-Serrano 1999).
At present this result is only significant at the
$\sim 2 \sigma$ level, due to the limited statistics.
However, it is certainly suggestive and
emphasises the potential use
of proximate DLAs as probes of the environment
of QSOs.

\item  Similarly, the number density of  PDLAs towards 
RLQs is $> 4$ times higher than that  of PDLAs towards RQQs.  
In fact, we find that $n(z)$ of PDLAs  towards RQQs is 
indistinguishable from that of intervening DLAs.
This result is consistent with the premise that RLQs preferentially
mark cluster environments.

\item   Since all of the PDLAs in the CORALS sample are found towards
optically bright QSOs ($ B < 20$) there appears to be no strong
extinction effect from dust.  Although previous
studies find that dust in \zae\ systems may redden the background
QSO (Carilli et al 1998), it appears that the amount of dust that
is present is modest and not sufficient to exclude a significant
number of quasars from optical surveys.

\item  We identify a new PDLA in the line of sight towards
B0405$-$331 which may exhibit \lya\ emission, based on residual
flux seen at the base of the saturated absorption trough.

\end{enumerate}

Lacy et al. (2001) have recently proposed a unified scheme
for radio-loud and radio-quiet QSOs
by showing that the radio luminosity of QSOs scales with the
black hole mass and the accretion rate onto the black hole.
Although the data support the broad increase in radio luminosity
with black hole mass, it is unclear whether accretion rate is
the physical driver of this relation (Dunlop et al. 2002).
We have presented in this paper evidence suggesting that radio-loud
QSOs are found preferentially in richer environments
than radio-quiet QSOs. At first sight, such a connection
between the pc scale of the central engine and the Mpc
scale of galaxy clusters may seem surprising. However,
it may simply be another manifestation of the relationship
uncovered by Magorrian et al. (1998) between black hole mass and 
bulge mass, if the most massive galaxies are also more likely
to be found in rich clusters.  

With the current limited statistics for PDLAs it is difficult
to extend the analysis of the DLA environment around QSOs.  As
the number of identified \zae\ systems increases 
(several candidates should be present in existing surveys) 
it will be possible to compare the
1-D QSO-DLA correlation function with the galaxy-QSO correlation
function in clusters (Mart\'{i}nez et al. 1999). Such a comparison 
will provide further clues to the nature of PDLAs, and in turn
contribute greatly to the 
investigation of QSO cluster environments.
Of course, we must bear in mind that the 3000 \kms\ cut-off
adopted as the definition for associated DLAs is somewhat arbitrary.
This point may be especially important since one of the CORALS PDLAs
is right on the 3000 \kms\ cut-off.
With improved statistics, it will be very interesting to investigate
$n(z)$ as a function of relative velocity from the QSO.  

\begin{acknowledgements}

We are grateful to C\'{e}line P\'{e}roux for communicating the statistics of
PDLAs from the survey of P\'{e}roux et al (2001) and to Dick Hunstead
for checking the SUMSS fields of some of these QSOs for their radio properties.
Thanks also are due to Marcin Sawicki and Gabriela Mallen-Ornelas for
continued stimulating discussions.

\end{acknowledgements}

\end{document}